\documentclass[letterpaper,compsoc,twoside]{IEEEtran}
\usepackage{fixltx2e} 
\usepackage{cmap} 
\usepackage{ifthen}
\usepackage[T1]{fontenc}
\usepackage[utf8]{inputenc}
\usepackage{amsmath}

\usepackage[font={small,it},labelfont=bf]{caption}
\usepackage{float}

\pdfoutput=1
\usepackage{scipy}
\makeatletter
\def\PY@reset{\let\PY@it=\relax \let\PY@bf=\relax%
    \let\PY@ul=\relax \let\PY@tc=\relax%
    \let\PY@bc=\relax \let\PY@ff=\relax}
\def\PY@tok#1{\csname PY@tok@#1\endcsname}
\def\PY@toks#1+{\ifx\relax#1\empty\else%
    \PY@tok{#1}\expandafter\PY@toks\fi}
\def\PY@do#1{\PY@bc{\PY@tc{\PY@ul{%
    \PY@it{\PY@bf{\PY@ff{#1}}}}}}}
\def\PY#1#2{\PY@reset\PY@toks#1+\relax+\PY@do{#2}}

\expandafter\def\csname PY@tok@gd\endcsname{\def\PY@tc##1{\textcolor[rgb]{0.63,0.00,0.00}{##1}}}
\expandafter\def\csname PY@tok@gu\endcsname{\let\PY@bf=\textbf\def\PY@tc##1{\textcolor[rgb]{0.50,0.00,0.50}{##1}}}
\expandafter\def\csname PY@tok@gt\endcsname{\def\PY@tc##1{\textcolor[rgb]{0.00,0.27,0.87}{##1}}}
\expandafter\def\csname PY@tok@gs\endcsname{\let\PY@bf=\textbf}
\expandafter\def\csname PY@tok@gr\endcsname{\def\PY@tc##1{\textcolor[rgb]{1.00,0.00,0.00}{##1}}}
\expandafter\def\csname PY@tok@cm\endcsname{\let\PY@it=\textit\def\PY@tc##1{\textcolor[rgb]{0.25,0.50,0.56}{##1}}}
\expandafter\def\csname PY@tok@vg\endcsname{\def\PY@tc##1{\textcolor[rgb]{0.73,0.38,0.84}{##1}}}
\expandafter\def\csname PY@tok@vi\endcsname{\def\PY@tc##1{\textcolor[rgb]{0.73,0.38,0.84}{##1}}}
\expandafter\def\csname PY@tok@mh\endcsname{\def\PY@tc##1{\textcolor[rgb]{0.13,0.50,0.31}{##1}}}
\expandafter\def\csname PY@tok@cs\endcsname{\def\PY@tc##1{\textcolor[rgb]{0.25,0.50,0.56}{##1}}\def\PY@bc##1{\setlength{\fboxsep}{0pt}\colorbox[rgb]{1.00,0.94,0.94}{\strut ##1}}}
\expandafter\def\csname PY@tok@ge\endcsname{\let\PY@it=\textit}
\expandafter\def\csname PY@tok@vc\endcsname{\def\PY@tc##1{\textcolor[rgb]{0.73,0.38,0.84}{##1}}}
\expandafter\def\csname PY@tok@il\endcsname{\def\PY@tc##1{\textcolor[rgb]{0.13,0.50,0.31}{##1}}}
\expandafter\def\csname PY@tok@go\endcsname{\def\PY@tc##1{\textcolor[rgb]{0.20,0.20,0.20}{##1}}}
\expandafter\def\csname PY@tok@cp\endcsname{\def\PY@tc##1{\textcolor[rgb]{0.00,0.44,0.13}{##1}}}
\expandafter\def\csname PY@tok@gi\endcsname{\def\PY@tc##1{\textcolor[rgb]{0.00,0.63,0.00}{##1}}}
\expandafter\def\csname PY@tok@gh\endcsname{\let\PY@bf=\textbf\def\PY@tc##1{\textcolor[rgb]{0.00,0.00,0.50}{##1}}}
\expandafter\def\csname PY@tok@ni\endcsname{\let\PY@bf=\textbf\def\PY@tc##1{\textcolor[rgb]{0.84,0.33,0.22}{##1}}}
\expandafter\def\csname PY@tok@nl\endcsname{\let\PY@bf=\textbf\def\PY@tc##1{\textcolor[rgb]{0.00,0.13,0.44}{##1}}}
\expandafter\def\csname PY@tok@nn\endcsname{\let\PY@bf=\textbf\def\PY@tc##1{\textcolor[rgb]{0.05,0.52,0.71}{##1}}}
\expandafter\def\csname PY@tok@no\endcsname{\def\PY@tc##1{\textcolor[rgb]{0.38,0.68,0.84}{##1}}}
\expandafter\def\csname PY@tok@na\endcsname{\def\PY@tc##1{\textcolor[rgb]{0.25,0.44,0.63}{##1}}}
\expandafter\def\csname PY@tok@nb\endcsname{\def\PY@tc##1{\textcolor[rgb]{0.00,0.44,0.13}{##1}}}
\expandafter\def\csname PY@tok@nc\endcsname{\let\PY@bf=\textbf\def\PY@tc##1{\textcolor[rgb]{0.05,0.52,0.71}{##1}}}
\expandafter\def\csname PY@tok@nd\endcsname{\let\PY@bf=\textbf\def\PY@tc##1{\textcolor[rgb]{0.33,0.33,0.33}{##1}}}
\expandafter\def\csname PY@tok@ne\endcsname{\def\PY@tc##1{\textcolor[rgb]{0.00,0.44,0.13}{##1}}}
\expandafter\def\csname PY@tok@nf\endcsname{\def\PY@tc##1{\textcolor[rgb]{0.02,0.16,0.49}{##1}}}
\expandafter\def\csname PY@tok@si\endcsname{\let\PY@it=\textit\def\PY@tc##1{\textcolor[rgb]{0.44,0.63,0.82}{##1}}}
\expandafter\def\csname PY@tok@s2\endcsname{\def\PY@tc##1{\textcolor[rgb]{0.25,0.44,0.63}{##1}}}
\expandafter\def\csname PY@tok@nt\endcsname{\let\PY@bf=\textbf\def\PY@tc##1{\textcolor[rgb]{0.02,0.16,0.45}{##1}}}
\expandafter\def\csname PY@tok@nv\endcsname{\def\PY@tc##1{\textcolor[rgb]{0.73,0.38,0.84}{##1}}}
\expandafter\def\csname PY@tok@s1\endcsname{\def\PY@tc##1{\textcolor[rgb]{0.25,0.44,0.63}{##1}}}
\expandafter\def\csname PY@tok@ch\endcsname{\let\PY@it=\textit\def\PY@tc##1{\textcolor[rgb]{0.25,0.50,0.56}{##1}}}
\expandafter\def\csname PY@tok@m\endcsname{\def\PY@tc##1{\textcolor[rgb]{0.13,0.50,0.31}{##1}}}
\expandafter\def\csname PY@tok@gp\endcsname{\let\PY@bf=\textbf\def\PY@tc##1{\textcolor[rgb]{0.78,0.36,0.04}{##1}}}
\expandafter\def\csname PY@tok@sh\endcsname{\def\PY@tc##1{\textcolor[rgb]{0.25,0.44,0.63}{##1}}}
\expandafter\def\csname PY@tok@ow\endcsname{\let\PY@bf=\textbf\def\PY@tc##1{\textcolor[rgb]{0.00,0.44,0.13}{##1}}}
\expandafter\def\csname PY@tok@sx\endcsname{\def\PY@tc##1{\textcolor[rgb]{0.78,0.36,0.04}{##1}}}
\expandafter\def\csname PY@tok@bp\endcsname{\def\PY@tc##1{\textcolor[rgb]{0.00,0.44,0.13}{##1}}}
\expandafter\def\csname PY@tok@c1\endcsname{\let\PY@it=\textit\def\PY@tc##1{\textcolor[rgb]{0.25,0.50,0.56}{##1}}}
\expandafter\def\csname PY@tok@o\endcsname{\def\PY@tc##1{\textcolor[rgb]{0.40,0.40,0.40}{##1}}}
\expandafter\def\csname PY@tok@kc\endcsname{\let\PY@bf=\textbf\def\PY@tc##1{\textcolor[rgb]{0.00,0.44,0.13}{##1}}}
\expandafter\def\csname PY@tok@c\endcsname{\let\PY@it=\textit\def\PY@tc##1{\textcolor[rgb]{0.25,0.50,0.56}{##1}}}
\expandafter\def\csname PY@tok@mf\endcsname{\def\PY@tc##1{\textcolor[rgb]{0.13,0.50,0.31}{##1}}}
\expandafter\def\csname PY@tok@err\endcsname{\def\PY@bc##1{\setlength{\fboxsep}{0pt}\fcolorbox[rgb]{1.00,0.00,0.00}{1,1,1}{\strut ##1}}}
\expandafter\def\csname PY@tok@mb\endcsname{\def\PY@tc##1{\textcolor[rgb]{0.13,0.50,0.31}{##1}}}
\expandafter\def\csname PY@tok@ss\endcsname{\def\PY@tc##1{\textcolor[rgb]{0.32,0.47,0.09}{##1}}}
\expandafter\def\csname PY@tok@sr\endcsname{\def\PY@tc##1{\textcolor[rgb]{0.14,0.33,0.53}{##1}}}
\expandafter\def\csname PY@tok@mo\endcsname{\def\PY@tc##1{\textcolor[rgb]{0.13,0.50,0.31}{##1}}}
\expandafter\def\csname PY@tok@kd\endcsname{\let\PY@bf=\textbf\def\PY@tc##1{\textcolor[rgb]{0.00,0.44,0.13}{##1}}}
\expandafter\def\csname PY@tok@mi\endcsname{\def\PY@tc##1{\textcolor[rgb]{0.13,0.50,0.31}{##1}}}
\expandafter\def\csname PY@tok@kn\endcsname{\let\PY@bf=\textbf\def\PY@tc##1{\textcolor[rgb]{0.00,0.44,0.13}{##1}}}
\expandafter\def\csname PY@tok@cpf\endcsname{\let\PY@it=\textit\def\PY@tc##1{\textcolor[rgb]{0.25,0.50,0.56}{##1}}}
\expandafter\def\csname PY@tok@kr\endcsname{\let\PY@bf=\textbf\def\PY@tc##1{\textcolor[rgb]{0.00,0.44,0.13}{##1}}}
\expandafter\def\csname PY@tok@s\endcsname{\def\PY@tc##1{\textcolor[rgb]{0.25,0.44,0.63}{##1}}}
\expandafter\def\csname PY@tok@kp\endcsname{\def\PY@tc##1{\textcolor[rgb]{0.00,0.44,0.13}{##1}}}
\expandafter\def\csname PY@tok@w\endcsname{\def\PY@tc##1{\textcolor[rgb]{0.73,0.73,0.73}{##1}}}
\expandafter\def\csname PY@tok@kt\endcsname{\def\PY@tc##1{\textcolor[rgb]{0.56,0.13,0.00}{##1}}}
\expandafter\def\csname PY@tok@sc\endcsname{\def\PY@tc##1{\textcolor[rgb]{0.25,0.44,0.63}{##1}}}
\expandafter\def\csname PY@tok@sb\endcsname{\def\PY@tc##1{\textcolor[rgb]{0.25,0.44,0.63}{##1}}}
\expandafter\def\csname PY@tok@k\endcsname{\let\PY@bf=\textbf\def\PY@tc##1{\textcolor[rgb]{0.00,0.44,0.13}{##1}}}
\expandafter\def\csname PY@tok@se\endcsname{\let\PY@bf=\textbf\def\PY@tc##1{\textcolor[rgb]{0.25,0.44,0.63}{##1}}}
\expandafter\def\csname PY@tok@sd\endcsname{\let\PY@it=\textit\def\PY@tc##1{\textcolor[rgb]{0.25,0.44,0.63}{##1}}}

\def\PYZus{\char`\_}
\def\PYZob{\char`\{}
\def\PYZcb{\char`\}}

\def\PYZsq{\char`\'}

\makeatother

\providecommand*{\DUrole}[2]{%
  \ifcsname DUrole#1\endcsname%
    \csname DUrole#1\endcsname{#2}%
  \else
    \ifcsname docutilsrole#1\endcsname%
      \csname docutilsrole#1\endcsname{#2}%
    \else%
      #2%
    \fi%
  \fi%
}

\ifthenelse{\isundefined{\hypersetup}}{
  \usepackage[colorlinks=true,linkcolor=blue,urlcolor=blue]{hyperref}
  \urlstyle{same} 
}{}

\begin{document}
\newcounter{footnotecounter}\title{PyCells for an Open Semiconductor Industry}\author{Sepideh Alassi$^{\setcounter{footnotecounter}{1}\fnsymbol{footnotecounter}\setcounter{footnotecounter}{2}\fnsymbol{footnotecounter}}$%
          \setcounter{footnotecounter}{1}\thanks{\fnsymbol{footnotecounter} %
          Corresponding author: \protect\href{mailto:sepideh.alassi@iis.fraunhofer.de}{sepideh.alassi@iis.fraunhofer.de}}\setcounter{footnotecounter}{2}\thanks{\fnsymbol{footnotecounter} Fraunhofer IIS}, Bertram Winter$^{\setcounter{footnotecounter}{3}\fnsymbol{footnotecounter}}$\setcounter{footnotecounter}{3}\thanks{\fnsymbol{footnotecounter} ams AG}\thanks{%

          \noindent%
          Copyright\,\copyright\,2015 Sepideh Alassi et al. This is an open-access article distributed under the terms of the Creative Commons Attribution License, which permits unrestricted use, distribution, and reproduction in any medium, provided the original author and source are credited. http://creativecommons.org/licenses/by/3.0/%
        }}\maketitle
          \renewcommand{\leftmark}{PROC. OF THE 8th EUR. CONF. ON PYTHON IN SCIENCE (EUROSCIPY 2015)}
          \renewcommand{\rightmark}{PYCELLS FOR AN OPEN SEMICONDUCTOR INDUSTRY}

\setcounter{page}{3}
\newcommand*{\docutilsroleref}{\ref}
\newcommand*{\docutilsrolelabel}{\label}
\AtEndDocument{\cleardoublepage}
\begin{abstract}In the modern semiconductor industry, automatic generation of parameterized and recurring layout structures plays an important role and should be present as a feature in Electronic Design Automation (EDA)-tools. Currently these layout generators are developed with a proprietary programming language and can be used with a specific EDA-tool. Therefore, the semiconductor companies find the development of the layout generators that can be used in all state of the art EDA-tools which support OpenAccess database appealing. The goal of this project is to develop computationally efficient layout generators with Python (PyCells), for ams AG technologies, that possess all the features of comprehensive layout generators.\end{abstract}

\section{Introduction%
  \label{introduction}%
}

The number of companies active in modern semiconductor business is increasing every day which raises the demand for cheaper EDA-tools. However, one of the main steps of designing an integrated circuit is drawing the physical structure, the layout. Thus the EDA-tools should have a feature for automatic generation of parametrized, recurring structures as pre-defined layouts in a library. The layout generators which are presently in use, are developed using a proprietary language, SKILL, and are only usable as a feature of a specific EDA-tool. On the other hand, the second major drawback of using the proprietary layout generators is the lack of interoperability characteristic. Consequently, the semiconductor manufacturers are not able to support their customers in an optimal way. Hence, the use of an open standard for an EDA database also allows chip designers to work on the same chip-design using tools, which often have more affordable prices, from different EDA vendors.

The aim of this project is developing interoperable layout generators with Python (PyCells), for H35, C35, and S35 technologies of ams AG, that can be used by any EDA-tool supporting OpenAccess database. The OpenAccess database will be further explained in the next section. For a certain set of parameters, PyCells should generate layouts identical to those generated by SKILL codes (PCells). Furthermore, extra interactive features such as stretching and auto-abutment are added to PyCells which are verified to be accurate and are optimized with respect to performance.

\section{Development%
  \label{development}%
}

OpenAccess \cite{OpenAccess} has been established as standard for storing design data in the semiconductor industry and builds the foundation for interoperability of EDA-tools. OpenAccess is a file based database and manages logical design data (schematics) as well as artwork data (layout) for manufacturing. The reference implementation for accessing design data within the database is written in C++. The company Ciranova (now acquired by Synopsys) developed a Python wrapper \cite{PythonAPI} for the OpenAccess C++ class library with the goal to access and modify design data through Python. Additionally, they provide an integrated development environment \cite{SynopsysPyCellStudio} that enables interactive development and debugging of Python code, while the effects on the design data are shown directly in a separate window of the GUI.

In our development, we have used Python API library to write Python codes which generate recurring layout structures based on a given set of parameters. Typical recurring structures are primitive devices that are used in integrated circuit design such as resistors, capacitors and transistors. The electrical properties of the mentioned devices are simplified to geometric dimensions (e. g. width and length) which serve as parameters for the Python scripts. The object oriented nature of Python matches very well with this task as shown in the following example:

A typical CMOS semiconductor process offers n-channel and p-channel transistors. Each of these transistor types can have multiple voltage ratings (e. g. 5V, 20V, 50V 120V), accordingly resulting in different layout structures. In general, transistors have similar gate, source and drain structures, yet they have different doping concentrations for the implant regions of the semiconductor material and different protection structures, which need to be reflected in the layout. Therefore, we have created a set of base classes in Python from which the more specific transistors inherit common structures, while adding only structures dedicated to the type of transistor. This approach reduces duplicate code and makes maintaining the scripts more intuitive when the number of devices increases. Finally the Python scripts are compiled to byte-code and stored inside the OpenAccess library. A plugin of the OpenAccess database ascertains loading the correct script when a parameter of the corresponding device is modified. In the next section, the development of PyCells is explained using a sample high voltage transistor.

\subsection{Sample PyCell Development%
  \label{sample-pycell-development}%
}

The process of developing a PyCell is described in this section step by step using the sample 20V MOS transistor, pmos20t. In order to simplify the development of new layout generators, the architecture shown in Figure \DUrole{ref}{architecture} has been established.\begin{figure}[]\noindent\makebox[\columnwidth][c]{\includegraphics[scale=0.30]{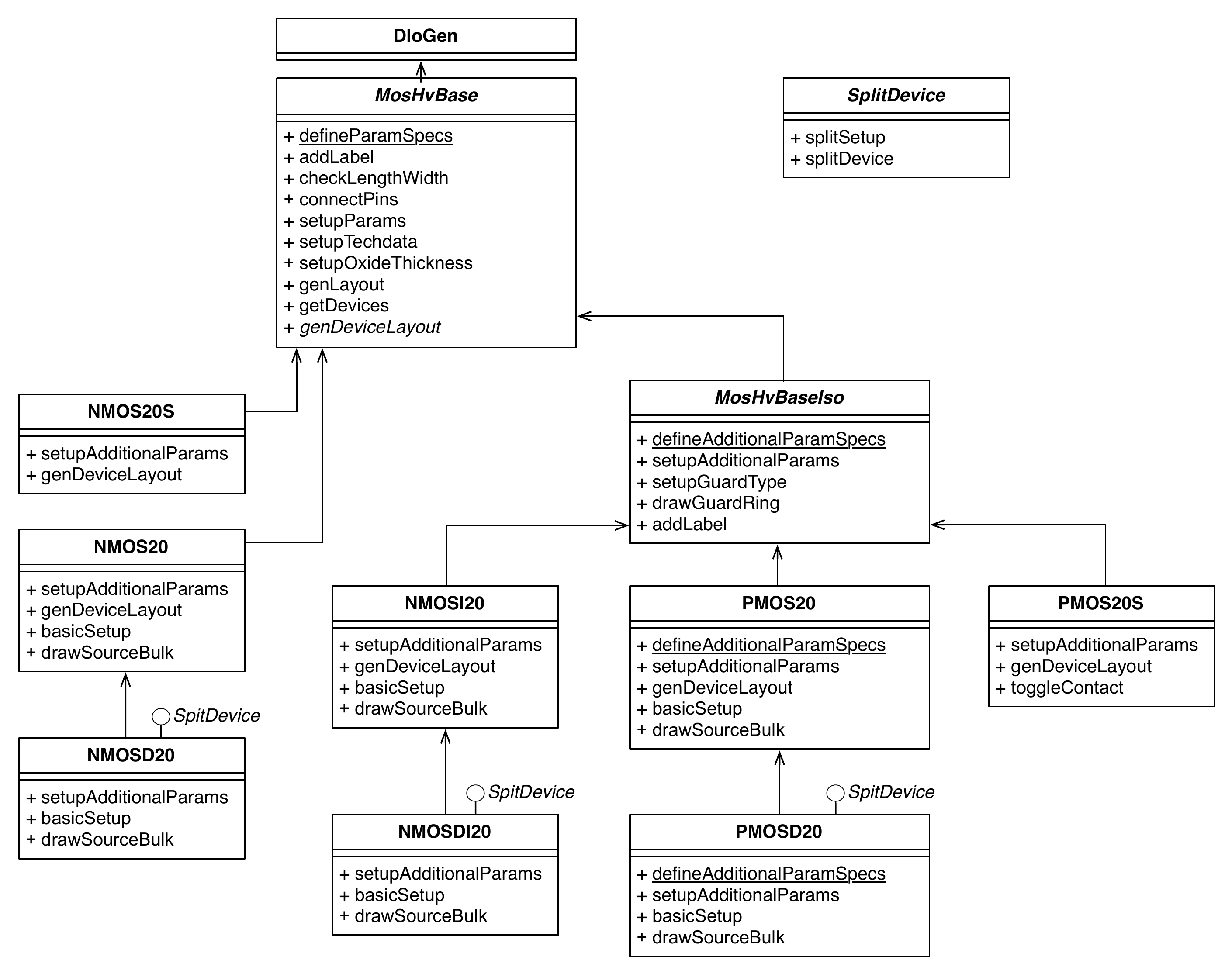}}
\caption{Class diagram for PyCells of 20V MOS transistors . \DUrole{label}{architecture}}
\end{figure}

For developing the PyCells, first step is setting the parameters of the transistor, such as length, width, number of fingers, type of guardring, etc. As shown in Figure \DUrole{ref}{param}\begin{figure}[]\noindent\makebox[\columnwidth][c]{\includegraphics[scale=0.60]{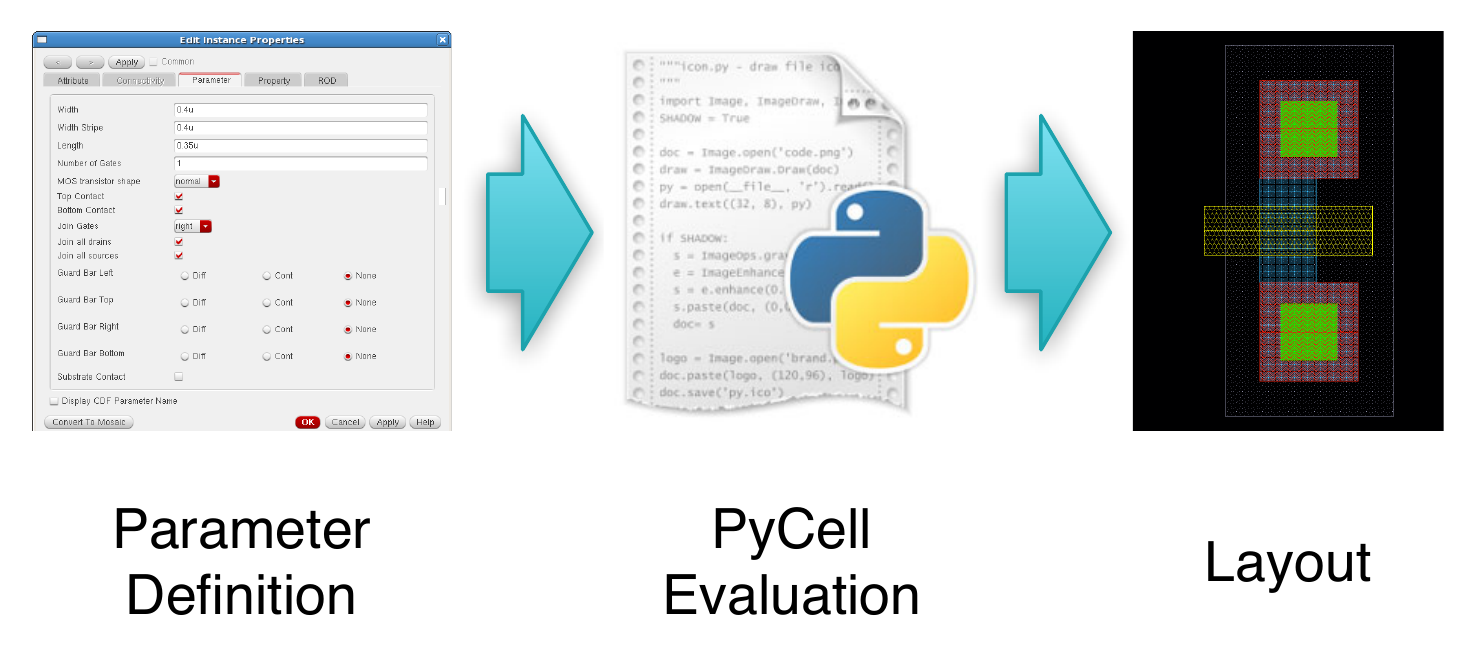}}
\caption{Evaluation flow of the parameters . \DUrole{label}{param}}
\end{figure}

The technology data (td) provided by foundry is read to have access to the layers and values to initialize some of the variables such as gate oxide thickness, etc. After the parameters are set and evaluated with respect to the acceptable thresholds defined by the foundry, the dimensions of layers as well as spaces between them are calculated.

After completing the setup step, the first structure to be generated is the polysilicon gate which has an octagonal shape drawn with a polysilicon layer (poly1) with a gap inside for drain contact. Later the gate structure is appended to the list of shapes which are connected to the gate pin.\begin{quotation}%
\begin{quote}
\begin{Verbatim}[commandchars=\\\{\},fontsize=\footnotesize]
\PY{n}{p1Gate} \PY{o}{=} \PY{n}{OctagonWithHole}\PY{p}{(}\PY{n}{layers}\PY{o}{.}\PY{n}{poly1}\PY{p}{,}
              \PY{n}{gateOuterBox}\PY{p}{,} \PY{n}{gateInnerBox}\PY{p}{,}
              \PY{n}{td}\PY{p}{[}\PY{l+s+s1}{\PYZsq{}}\PY{l+s+s1}{p1\PYZus{}corn}\PY{l+s+s1}{\PYZsq{}}\PY{p}{]}\PY{p}{,} \PY{l+m+mi}{0}\PY{p}{)}
\PY{n+nb+bp}{self}\PY{o}{.}\PY{n}{pins}\PY{p}{[}\PY{l+s+s1}{\PYZsq{}}\PY{l+s+s1}{G}\PY{l+s+s1}{\PYZsq{}}\PY{p}{]}\PY{o}{.}\PY{n}{append}\PY{p}{(}\PY{n}{p1Gate}\PY{p}{)}
\end{Verbatim}

\end{quote}
\end{quotation}Other structures of the transistor, such as diffusion layer, drain and source pins and the contacts of them, are generated serially in a predefined order. (see Figure \DUrole{ref}{pin})\begin{figure}[]\noindent\makebox[\columnwidth][c]{\includegraphics[scale=0.50]{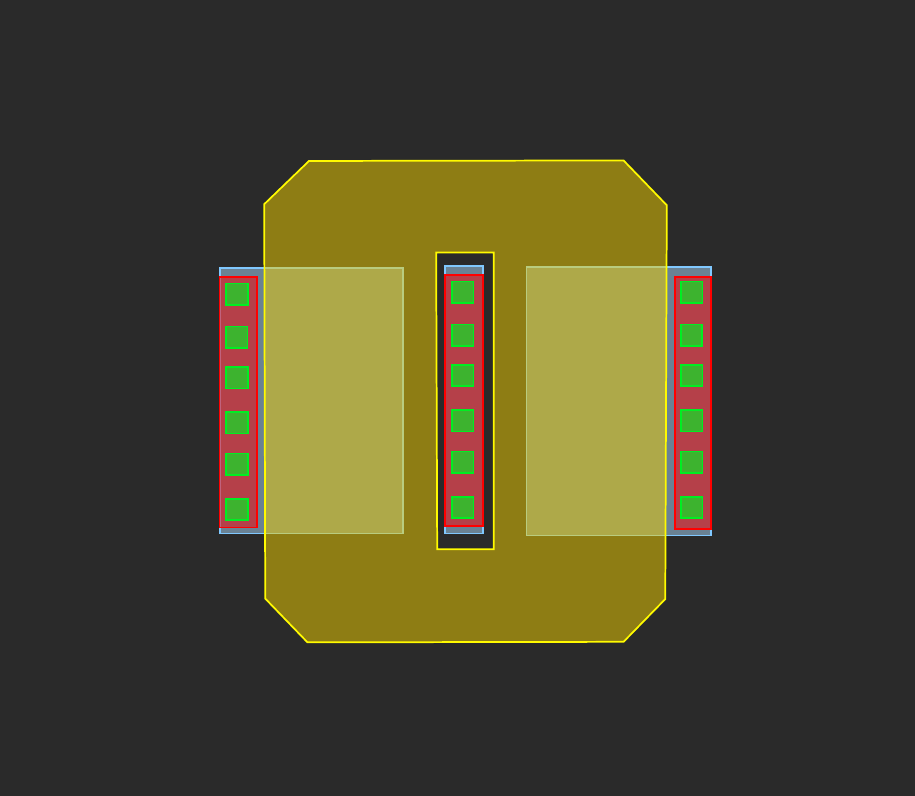}}
\caption{Creation of the source and drain contacts. \DUrole{label}{pin}}
\end{figure}

In order to increase the breakdown voltage of the device, regions with different doping concentrations have to be created at the drain terminal of the transistor. (see Figure \DUrole{ref}{well})\begin{figure}[]\noindent\makebox[\columnwidth][c]{\includegraphics[scale=0.50]{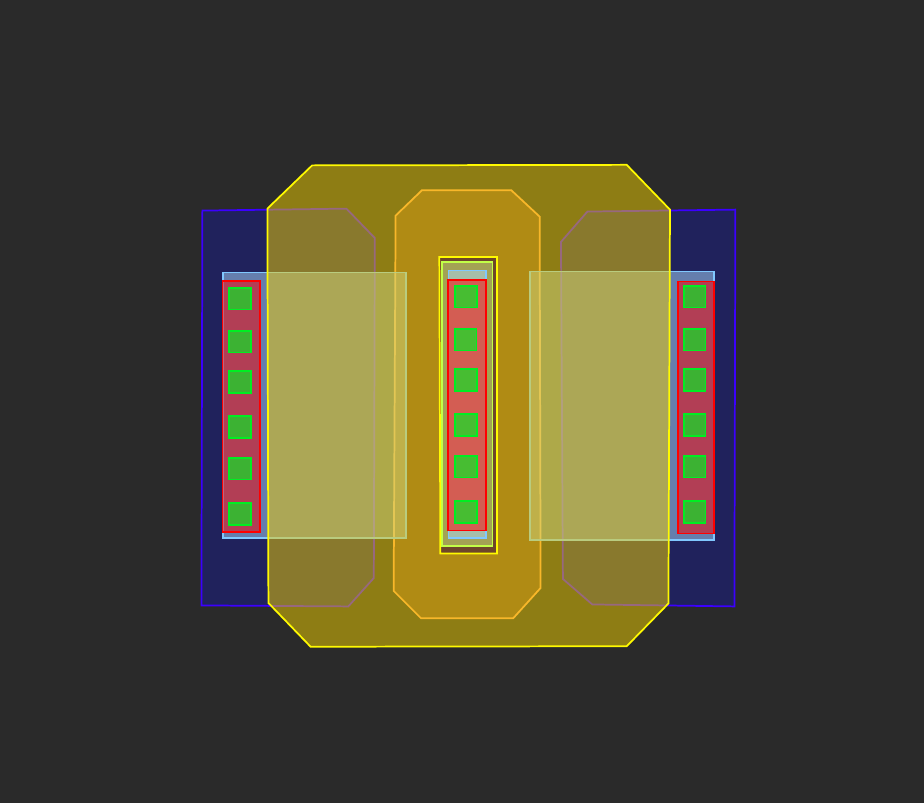}}
\caption{Deep and shallow p-well under the drain. \DUrole{label}{well}}
\end{figure}

Similarly an n-well should be added to define the device body, followed by a p-implant layer on source and drain. Furthermore, the bulk connection of the pmos20t device is accomplished by a contact ring around the device which has to make the connection to the n-type isolation well, for which n-type implant and a shallow n-well are required. (see Figure \DUrole{ref}{bulk})\begin{figure}[]\noindent\makebox[\columnwidth][c]{\includegraphics[scale=0.50]{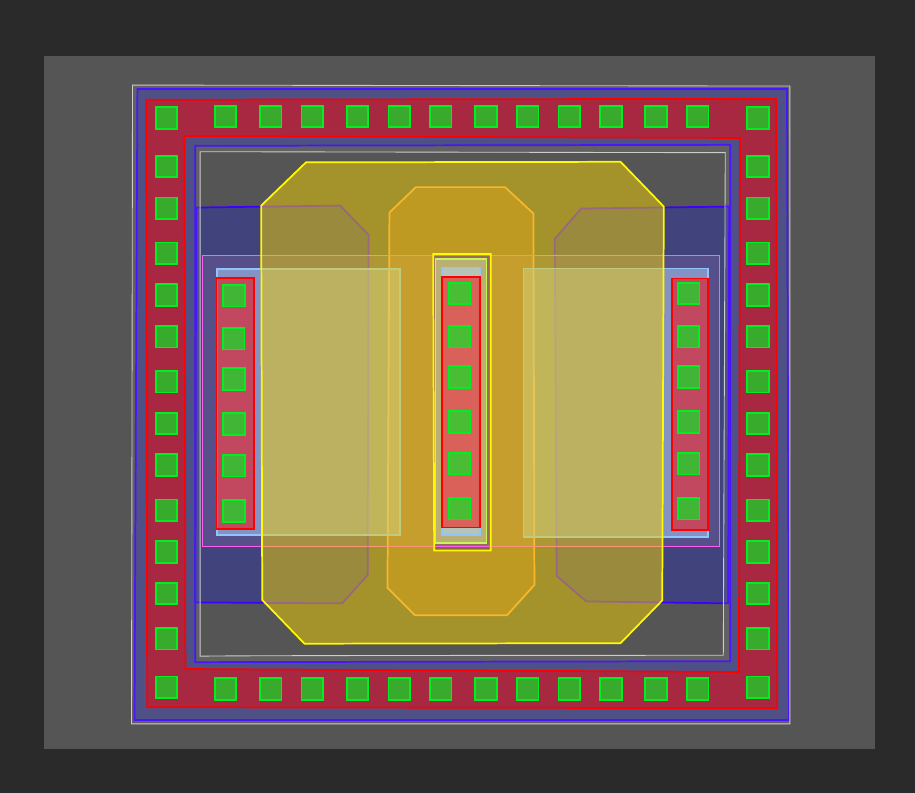}}
\caption{Bulk contact as a ring with an n-implant and an n-well. \DUrole{label}{bulk}}
\end{figure}

Lastly, the overall device structures, such as definition layer for the thickness of the gate oxide, should be added to PyCell. The resulting layout for pmos20t is shown in Figure \DUrole{ref}{pmos20}.\begin{figure}[]\noindent\makebox[\columnwidth][c]{\includegraphics[scale=0.40]{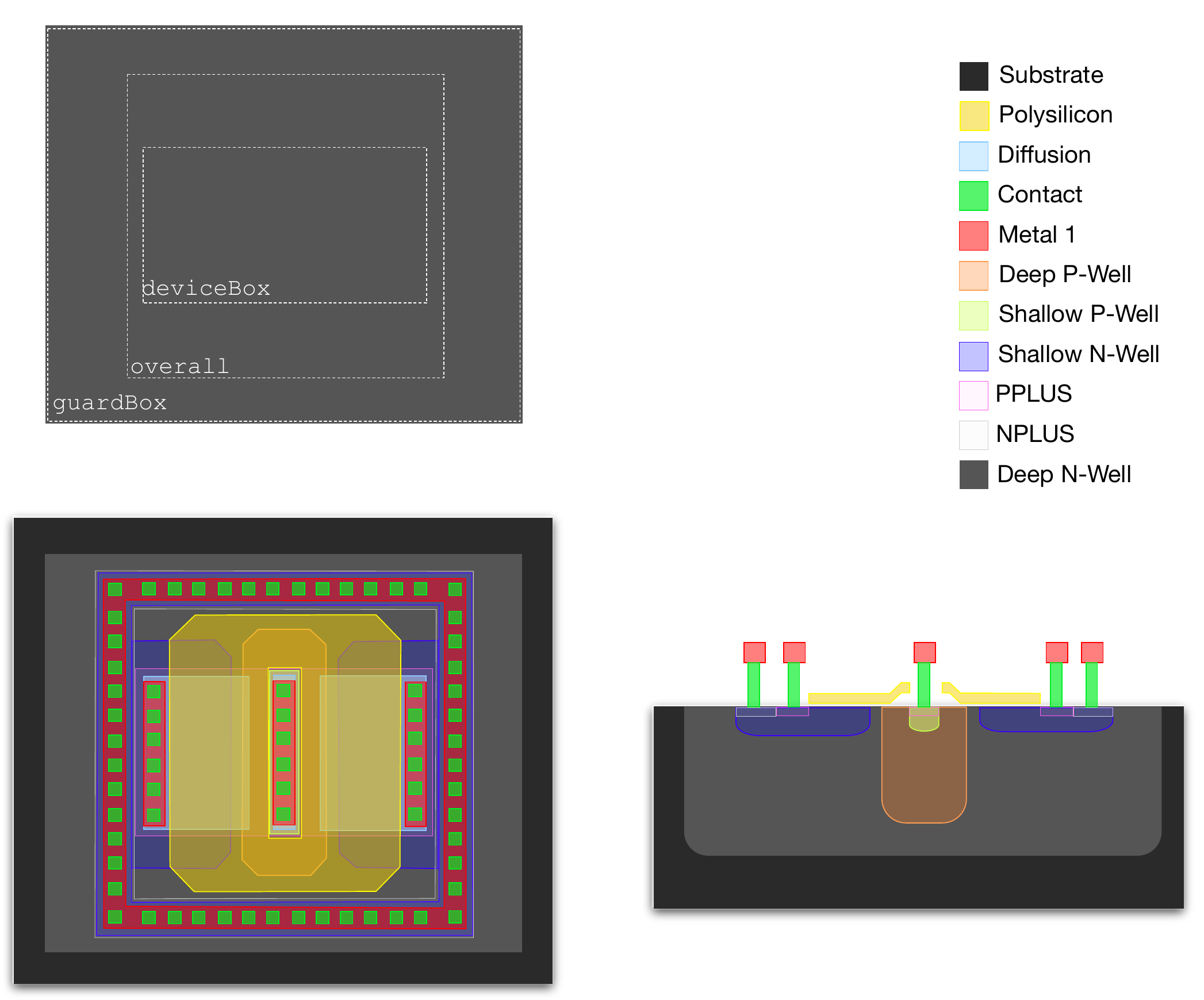}}
\caption{Final layout of a pmos20t transistor. \DUrole{label}{pmos20}}
\end{figure}

Figure \DUrole{ref}{pmos50} shows the layout generated for a transistor with 50V guard ring and multiplication factor of 2.\begin{figure}[]\noindent\makebox[\columnwidth][c]{\includegraphics[scale=0.35]{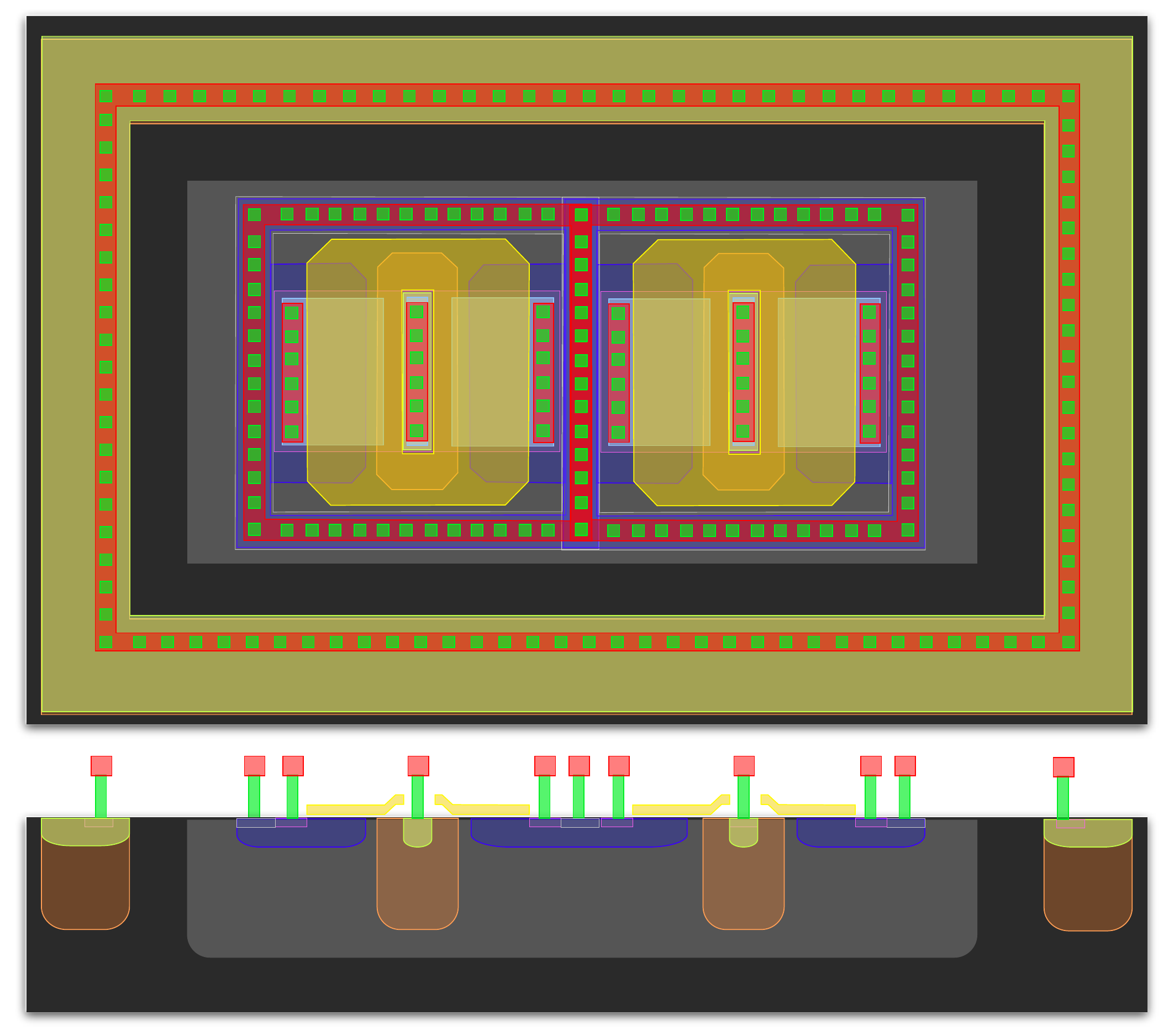}}
\caption{Layout with a 50V guard ring and a device multiplier of 2. \DUrole{label}{pmos50}}
\end{figure}

\subsection{Optimization%
  \label{optimization}%
}

The computation time plays a crucial role in designing integrated circuits. The complexity of the designs and the amount of operations necessary to generate layouts of high voltage devices or the large devices increase the computation time significantly. For this reason, throughout this project the PyCells are developed to be in the most optimized shape possible without losing the clarity of the structure of the codes. For optimizing the codes, data aggregation and optimal use of local variables as well as iterations are taken into account. This has resulted in faster execution of the PyCell codes for generating large and complex devices in comparison with execution time of PCells for identical devices.

\section{Stretching%
  \label{stretching}%
}

Along with the basic features such as specifying parameters for recurring structures, the Python API also offers more enhanced interactive features with the EDA-tool \cite{PythonAPI}. Stretching, which is one of these features allows changing parameters of an instance interactively. This helps the engineer to fit instances of an analog design into the available space, without affecting the electrical properties of the instance. For example, the capacitance value of a capacitor depends to a large extent on its area. Therefore, the shape can be varied while maintaining a constant area. An example of stretching is shown for a rectangular capacitor in Figure \DUrole{ref}{capfig}:\begin{equation}
\label{capequ}
C \propto W \cdot L
\end{equation}\begin{figure}[]\noindent\makebox[\columnwidth][c]{\includegraphics[scale=0.20]{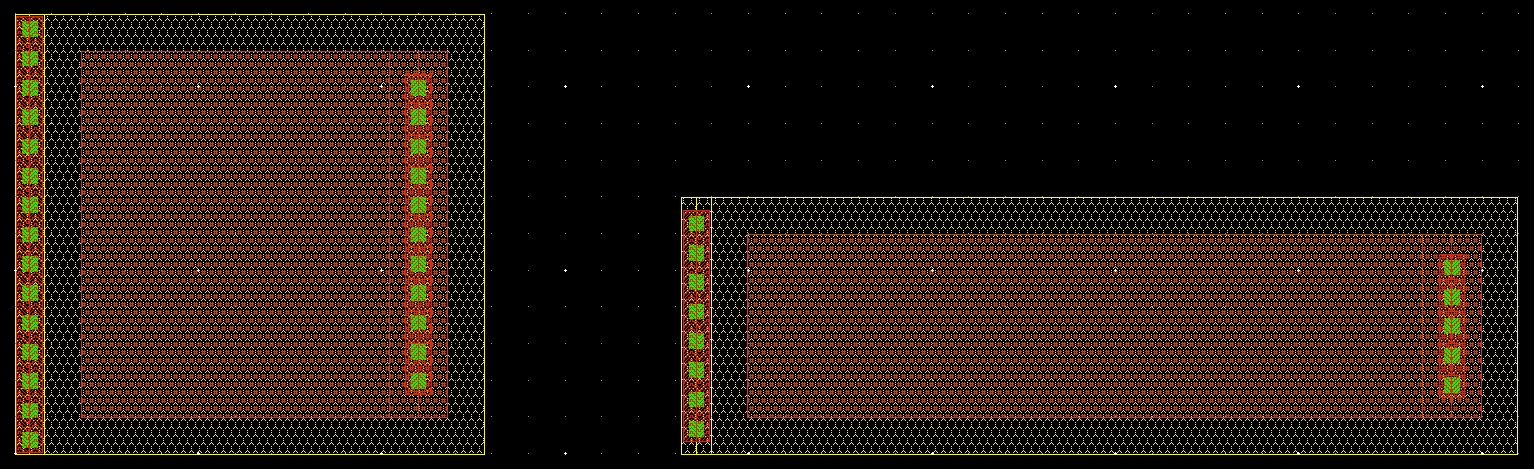}}
\caption{Stretching of a capacitor by changing L and W ratio, but keeping its area constant. \DUrole{label}{capfig}}
\end{figure}The electrical properties of resistors and transistors are proportional to the ratio of length and width:\begin{equation}
\label{resequ}
R \propto \frac{L}{W}
\end{equation}\begin{equation}
\label{tranequ}
g_m \propto \frac{W}{L}
\end{equation}Therefore, one cannot simply fit them into a given shape without affecting the behavior of the design. One option is to allow the resistors to have bends (as shown in Figure \DUrole{ref}{resfig}) or to change the number of fingers of transistors (see Figure \DUrole{ref}{transfig}). Stretching the gate of a transistor to change its dimensions is also possible (see Figure \DUrole{ref}{stretchgate}).\begin{figure}[]\noindent\makebox[\columnwidth][c]{\includegraphics[scale=0.20]{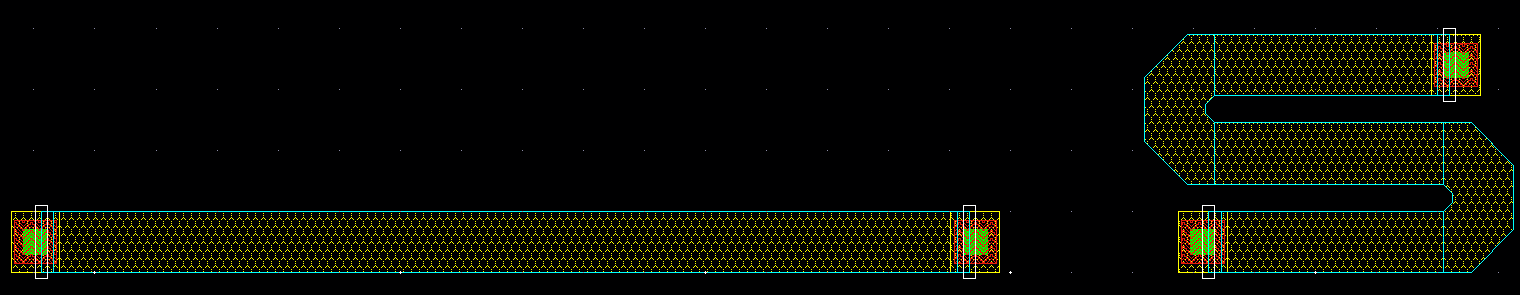}}
\caption{Stretching of a resistor by bending the device in a snake structure, but keeping length and width constant. \DUrole{label}{resfig}}
\end{figure}\begin{figure}[]\noindent\makebox[\columnwidth][c]{\includegraphics[scale=0.40]{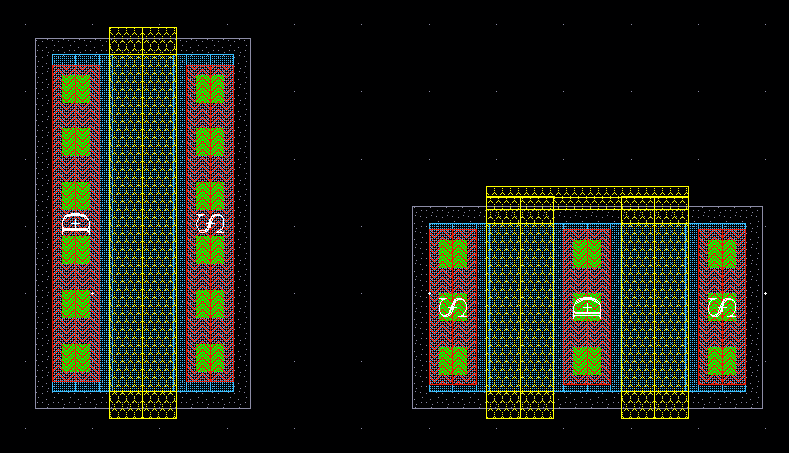}}
\caption{Stretching of a transistor by increasing the number of fingers, but keeping the gate to source/drain edge constant. \DUrole{label}{transfig}}
\end{figure}\begin{figure}[]\noindent\makebox[\columnwidth][c]{\includegraphics[scale=0.80]{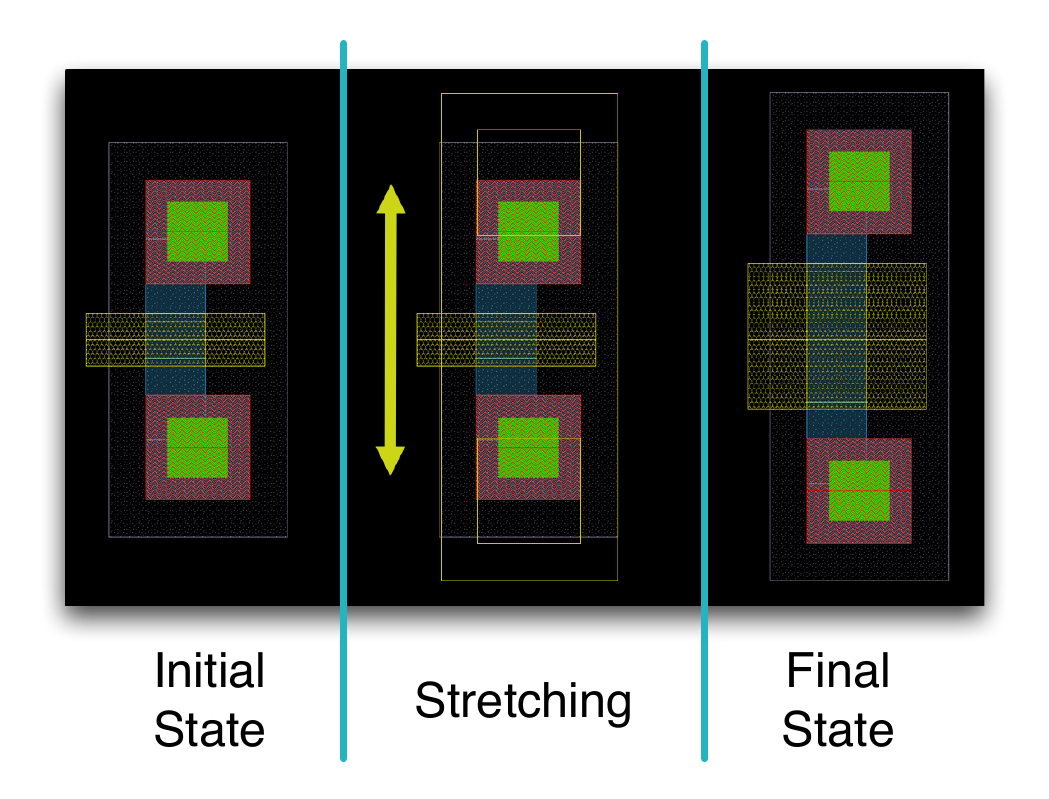}}
\caption{Stretching the gate of a transistor \DUrole{label}{stretchgate}}
\end{figure}

The stretch handles appear as diamond shaped elements on the GUI, indicating the place where the user can click and drag the object in order to change its dimensions. These handles are stored as properties in the OpenAccess database and are defined within the Python source code. The stretch handles have to be associated to the shape and parameters of the layout. Stretching can be limited to minimum and maximum boundaries on both horizontal and vertical directions. Snap resolution can be also be specified in definition of stretch handles, an example of which is shown below:\begin{Verbatim}[commandchars=\\\{\},fontsize=\footnotesize]
\PY{n}{stretchHandle}\PY{p}{(}
    \PY{n}{name} \PY{o}{=} \PY{l+s+s1}{\PYZsq{}}\PY{l+s+s1}{width\PYZus{}handle\PYZus{}left}\PY{l+s+s1}{\PYZsq{}}\PY{p}{,}
    \PY{n}{shape} \PY{o}{=} \PY{n}{poly1GateShape}\PY{p}{,}
        \PY{n}{parameter} \PY{o}{=} \PY{l+s+s1}{\PYZsq{}}\PY{l+s+s1}{wtot}\PY{l+s+s1}{\PYZsq{}}\PY{p}{,}
        \PY{n}{location} \PY{o}{=} \PY{n}{CENTER\PYZus{}LEFT}\PY{p}{,}
        \PY{n}{direction} \PY{o}{=} \PY{n}{EAST\PYZus{}WEST}\PY{p}{,}
        \PY{n}{stretchType} \PY{o}{=} \PY{l+s+s1}{\PYZsq{}}\PY{l+s+s1}{relative}\PY{l+s+s1}{\PYZsq{}}\PY{p}{,}
        \PY{n}{minVal} \PY{o}{=} \PY{l+m+mf}{0.4}\PY{p}{,}
        \PY{n}{maxVal} \PY{o}{=} \PY{l+m+mf}{10000.0}
\PY{p}{)}
\end{Verbatim}
The GUI of EDA-tool displays the stretch handles as small diamonds on layout, which can be dragged by the engineer graphically to change the value of parameters using mouse. By releasing the stretch handle, the Python code is automatically invoked and the layout structure is adapted according to the new values of parameters.

\section{Abutment%
  \label{abutment}%
}

Another advanced feature is auto-abutment which is used to make the layout more compact. In case of placing two instances next to each other, the layout can be adapted in a way that common structures are shared between the instances. This minimizes the layout area and two instances appear to be merged. Similar to stretching, auto-abutment is defined by additional properties associated with shapes in the OpenAccess database. In PyCells, abutment is defined for a graphical structure by the autoAbutment() function:\begin{Verbatim}[commandchars=\\\{\},fontsize=\footnotesize]
\PY{n}{autoAbutment}\PY{p}{(}
    \PY{n}{shape} \PY{o}{=} \PY{n}{drain}\PY{p}{,}
    \PY{n}{pinSize} \PY{o}{=} \PY{n+nb+bp}{self}\PY{o}{.}\PY{n}{w}\PY{p}{,}
    \PY{n}{directions} \PY{o}{=} \PY{p}{[}\PY{n}{WEST}\PY{p}{]}\PY{p}{,}
    \PY{n}{abutClass} \PY{o}{=} \PY{l+s+s1}{\PYZsq{}}\PY{l+s+s1}{mos\PYZus{}drain}\PY{l+s+s1}{\PYZsq{}}\PY{p}{,}
    \PY{n}{abut2PinEqual} \PY{o}{=} \PY{p}{[}\PY{p}{\PYZob{}}\PY{l+s+s1}{\PYZsq{}}\PY{l+s+s1}{spacing}\PY{l+s+s1}{\PYZsq{}}\PY{p}{:} \PY{l+m+mf}{0.0}\PY{p}{\PYZcb{}}\PY{p}{,}
        \PY{p}{\PYZob{}}\PY{l+s+s1}{\PYZsq{}}\PY{l+s+s1}{diffLeftStyle}\PY{l+s+s1}{\PYZsq{}}\PY{p}{:} \PY{l+s+s1}{\PYZsq{}}\PY{l+s+s1}{DiffHalf}\PY{l+s+s1}{\PYZsq{}}\PY{p}{\PYZcb{}}\PY{p}{,}
        \PY{p}{\PYZob{}}\PY{l+s+s1}{\PYZsq{}}\PY{l+s+s1}{diffLeftStyle}\PY{l+s+s1}{\PYZsq{}}\PY{p}{:} \PY{l+s+s1}{\PYZsq{}}\PY{l+s+s1}{DiffHalf}\PY{l+s+s1}{\PYZsq{}}\PY{p}{\PYZcb{}}\PY{p}{]}\PY{p}{,}
    \PY{n}{abut2PinBigger} \PY{o}{=} \PY{p}{[}\PY{p}{\PYZob{}}\PY{l+s+s1}{\PYZsq{}}\PY{l+s+s1}{spacing}\PY{l+s+s1}{\PYZsq{}}\PY{p}{:} \PY{l+m+mf}{0.0}\PY{p}{\PYZcb{}}\PY{p}{,}
        \PY{p}{\PYZob{}}\PY{l+s+s1}{\PYZsq{}}\PY{l+s+s1}{diffLeftStyle}\PY{l+s+s1}{\PYZsq{}}\PY{p}{:} \PY{l+s+s1}{\PYZsq{}}\PY{l+s+s1}{DiffEdgeAbut}\PY{l+s+s1}{\PYZsq{}}\PY{p}{\PYZcb{}}\PY{p}{,}
        \PY{p}{\PYZob{}}\PY{l+s+s1}{\PYZsq{}}\PY{l+s+s1}{diffLeftStyle}\PY{l+s+s1}{\PYZsq{}}\PY{p}{:} \PY{l+s+s1}{\PYZsq{}}\PY{l+s+s1}{DiffEdgeAbut}\PY{l+s+s1}{\PYZsq{}}\PY{p}{\PYZcb{}}\PY{p}{]}\PY{p}{,}
    \PY{n}{abut3PinBigger} \PY{o}{=} \PY{p}{[}\PY{p}{\PYZob{}}\PY{l+s+s1}{\PYZsq{}}\PY{l+s+s1}{spacing}\PY{l+s+s1}{\PYZsq{}}\PY{p}{:} \PY{l+m+mf}{0.0}\PY{p}{\PYZcb{}}\PY{p}{,}
        \PY{p}{\PYZob{}}\PY{l+s+s1}{\PYZsq{}}\PY{l+s+s1}{diffLeftStyle}\PY{l+s+s1}{\PYZsq{}}\PY{p}{:} \PY{l+s+s1}{\PYZsq{}}\PY{l+s+s1}{ContactEdgeAbut2}\PY{l+s+s1}{\PYZsq{}}\PY{p}{\PYZcb{}}\PY{p}{,}
        \PY{p}{\PYZob{}}\PY{l+s+s1}{\PYZsq{}}\PY{l+s+s1}{diffLeftStyle}\PY{l+s+s1}{\PYZsq{}}\PY{p}{:} \PY{l+s+s1}{\PYZsq{}}\PY{l+s+s1}{ContactEdgeAbut2}\PY{l+s+s1}{\PYZsq{}}\PY{p}{\PYZcb{}}\PY{p}{]}\PY{p}{,}
    \PY{n}{abut3PinEqual} \PY{o}{=} \PY{p}{[}\PY{p}{\PYZob{}}\PY{l+s+s1}{\PYZsq{}}\PY{l+s+s1}{spacing}\PY{l+s+s1}{\PYZsq{}}\PY{p}{:} \PY{l+m+mf}{0.0}\PY{p}{\PYZcb{}}\PY{p}{,}
        \PY{p}{\PYZob{}}\PY{l+s+s1}{\PYZsq{}}\PY{l+s+s1}{diffLeftStyle}\PY{l+s+s1}{\PYZsq{}}\PY{p}{:} \PY{l+s+s1}{\PYZsq{}}\PY{l+s+s1}{DiffAbut}\PY{l+s+s1}{\PYZsq{}}\PY{p}{\PYZcb{}}\PY{p}{,}
        \PY{p}{\PYZob{}}\PY{l+s+s1}{\PYZsq{}}\PY{l+s+s1}{diffLeftStyle}\PY{l+s+s1}{\PYZsq{}}\PY{p}{:} \PY{l+s+s1}{\PYZsq{}}\PY{l+s+s1}{ContactEdgeAbut2}\PY{l+s+s1}{\PYZsq{}}\PY{p}{\PYZcb{}}\PY{p}{]}\PY{p}{,}
    \PY{n}{abut2PinSmaller} \PY{o}{=} \PY{p}{[}\PY{p}{\PYZob{}}\PY{l+s+s1}{\PYZsq{}}\PY{l+s+s1}{spacing}\PY{l+s+s1}{\PYZsq{}}\PY{p}{:} \PY{l+m+mf}{0.0}\PY{p}{\PYZcb{}}\PY{p}{,}
        \PY{p}{\PYZob{}}\PY{l+s+s1}{\PYZsq{}}\PY{l+s+s1}{diffLeftStyle}\PY{l+s+s1}{\PYZsq{}}\PY{p}{:} \PY{l+s+s1}{\PYZsq{}}\PY{l+s+s1}{DiffEdgeAbut}\PY{l+s+s1}{\PYZsq{}}\PY{p}{\PYZcb{}}\PY{p}{,}
        \PY{p}{\PYZob{}}\PY{l+s+s1}{\PYZsq{}}\PY{l+s+s1}{diffLeftStyle}\PY{l+s+s1}{\PYZsq{}}\PY{p}{:} \PY{l+s+s1}{\PYZsq{}}\PY{l+s+s1}{DiffEdgeAbut}\PY{l+s+s1}{\PYZsq{}}\PY{p}{\PYZcb{}}\PY{p}{]}\PY{p}{,}
    \PY{n}{abut3PinSmaller} \PY{o}{=} \PY{p}{[}\PY{p}{\PYZob{}}\PY{l+s+s1}{\PYZsq{}}\PY{l+s+s1}{spacing}\PY{l+s+s1}{\PYZsq{}}\PY{p}{:} \PY{l+m+mf}{0.0}\PY{p}{\PYZcb{}}\PY{p}{,}
        \PY{p}{\PYZob{}}\PY{l+s+s1}{\PYZsq{}}\PY{l+s+s1}{diffLeftStyle}\PY{l+s+s1}{\PYZsq{}}\PY{p}{:} \PY{l+s+s1}{\PYZsq{}}\PY{l+s+s1}{DiffEdgeAbut}\PY{l+s+s1}{\PYZsq{}}\PY{p}{\PYZcb{}}\PY{p}{,}
        \PY{p}{\PYZob{}}\PY{l+s+s1}{\PYZsq{}}\PY{l+s+s1}{diffLeftStyle}\PY{l+s+s1}{\PYZsq{}}\PY{p}{:} \PY{l+s+s1}{\PYZsq{}}\PY{l+s+s1}{DiffEdgeAbut}\PY{l+s+s1}{\PYZsq{}}\PY{p}{\PYZcb{}}\PY{p}{]}\PY{p}{,}
    \PY{n}{noAbut} \PY{o}{=} \PY{p}{[}\PY{p}{\PYZob{}}\PY{l+s+s1}{\PYZsq{}}\PY{l+s+s1}{spacing}\PY{l+s+s1}{\PYZsq{}}\PY{p}{:} \PY{l+m+mf}{0.4} \PY{p}{\PYZcb{}}\PY{p}{]}
\PY{p}{)}
\end{Verbatim}
This function provides a variety of parameters to define which attributes are compatible with abutment. Only instances that have common layout structures can be merged. Furthermore, there are arguments of this function to specify different types of abutment, for example in cases where the instances have common structures but different dimensions. It is also important to take logical information of the design into account, as it is only allowed to merge structures which are logically connected (i.e. having the same net). The resulting layout can be further diminished, when the two instances are the only devices connected to the net. In this case, the structures between the two instances, that would allow connections to the net, can be omitted. The abutment process is shown in Figure \DUrole{ref}{abutment}, and Figure \DUrole{ref}{abutfig} depicts different cases of abutment.\begin{figure}[]\noindent\makebox[\columnwidth][c]{\includegraphics[scale=0.80]{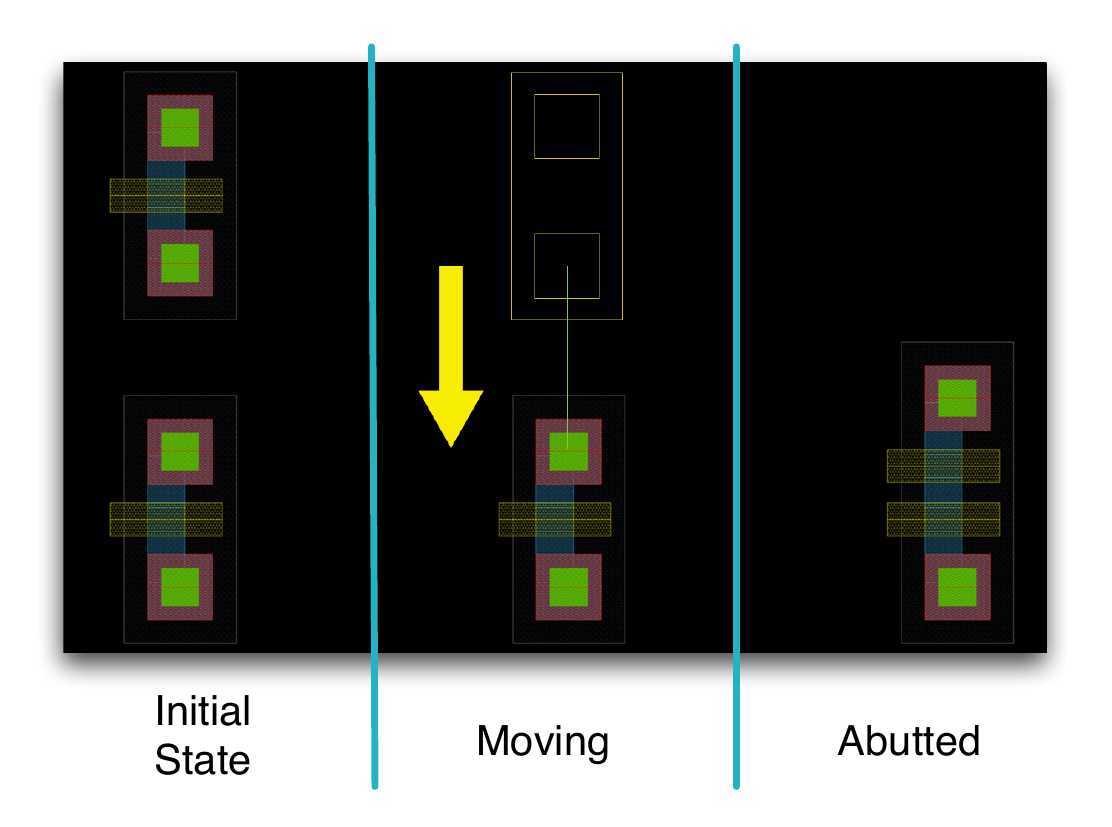}}
\caption{Abutment of two transistors \DUrole{label}{abutment}}
\end{figure}\begin{figure}[]\noindent\makebox[\columnwidth][c]{\includegraphics[scale=0.80]{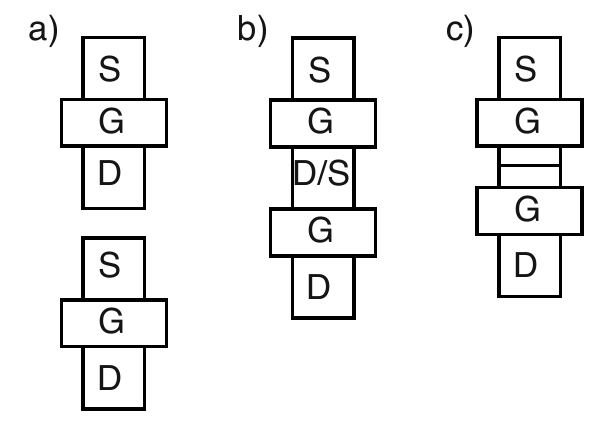}}
\caption{Abutment of two transistors: a) no abutment, as the gap between transistors is too large b) transistors abutted, but leave source/drain pin to connect to another instance c) transistors abutted, no connection to source/drain needed \DUrole{label}{abutfig}}
\end{figure}

In order to abut two transistors, the layout designer should drag one instance and place it such that the drain contacts overlap. Consequently the abutment feature of the PyCell will be triggered and two instances will be merged. If one instance is relocated, each of the other instances will get its initial structure.

\section{Verification%
  \label{verification}%
}

The last part of the project which is of utmost importance is verifying the PyCells. We have considered four processes in order to verify the accuracy of PyCells: Design Rule Check (DRC), Layout Versus Schematic (LVS), Schematic Driven Layout (SDL), and database Comparison (DBCOMP), as explained below.

\subsection{Design Rule Check (DRC)%
  \label{design-rule-check-drc}%
}

The semiconductor manufacturers define a set of design rules for every technology with which the layout of every design must comply. Therefore, all the layouts generated by PyCells with various parameters, are checked for design rule violations in order to ensure the design accuracy of them. A DRC verification software checks the layouts and highlights the violations of design rules, such as insufficient space between layers, overlapping layers, incorrect dimension of layers, etc. Some of these design rules are shown in Figure \DUrole{ref}{DRC}.\begin{figure}[]\noindent\makebox[\columnwidth][c]{\includegraphics[scale=0.50]{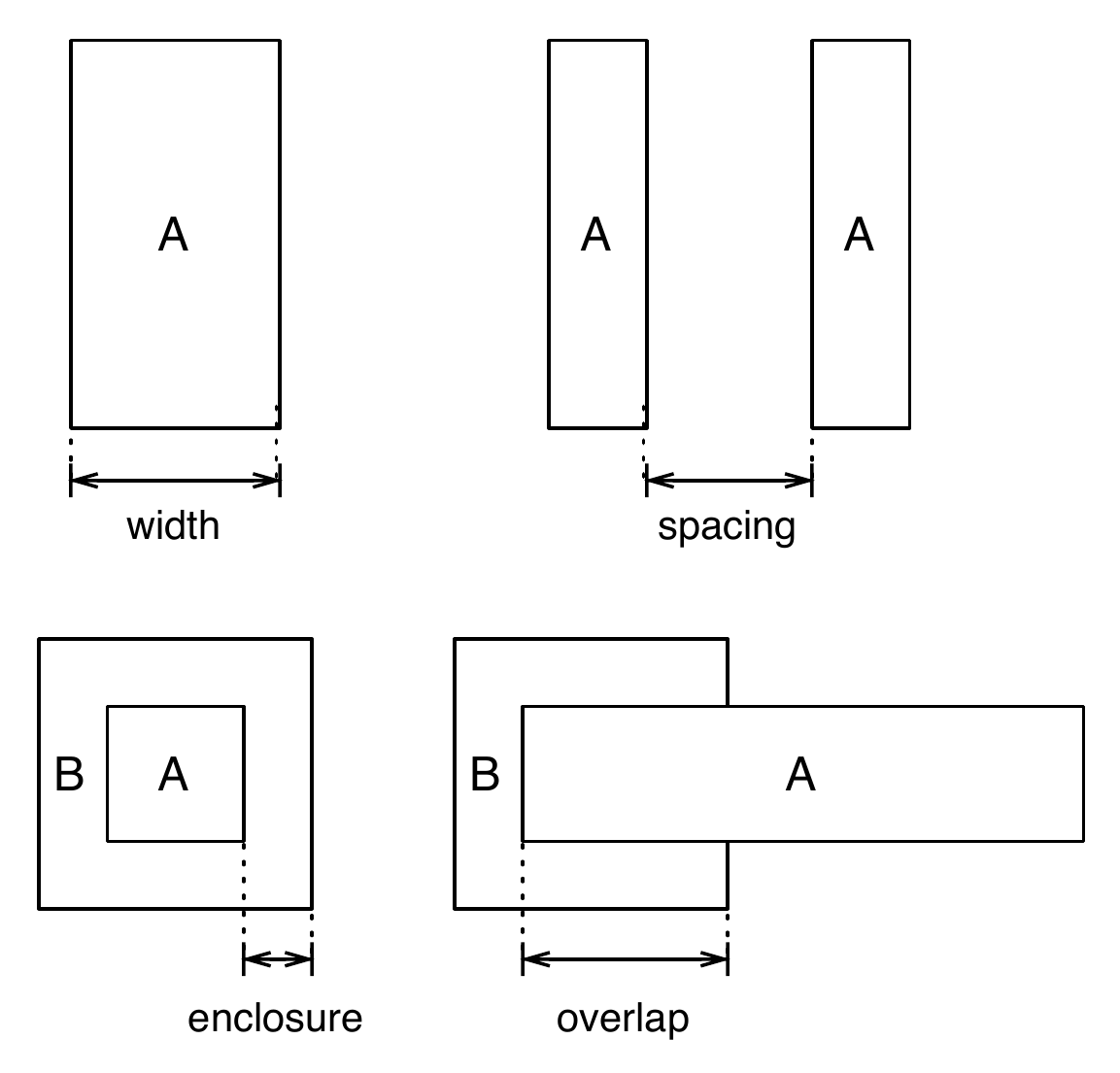}}
\caption{Design rules check \DUrole{label}{DRC}}
\end{figure}

\subsection{Layout versus Schematic (LVS)%
  \label{layout-versus-schematic-lvs}%
}

The layout of a design should match with its schematic with respect to type, number of devices, connections and topological parameters. In this verification process, we ensure that the layout of a sample design which is generated by PyCells matches perfectly with its corresponding schematic. The LVS tool, first extracts the netlists of layout and schematic of a design according to design rules defined by semiconductor manufacturer. In these netlists, the devices used in design and their connections as well as topological parameters (such as area, perimeter, etc) are listed. The LVS tool compares these netlists and reports if the schematic and layout of the design match completely with respect to type and connection of devices as well as topology of them. It also reports the existence of fragmented nodes which appear if the connections are not identical in the front- and back-end of the design.

\subsection{Schematic Driven Layout (SDL)%
  \label{schematic-driven-layout-sdl}%
}

PyCell codes must include the definition of pins regarding type of connections, weak connection (by polysilicon) or strong connection (by metal), and also the definition of shapes requiring external connections. Having types of connections implemented in PyCells, the SDL Navigator tool can attain the schematic of a sample design and generate the layout of it by calling PyCells with respective parameters. After generation of the layout, SDL Navigator checks the connectivity features using flylines (the wire-like shapes which appear on layout and show the connections between the ports).

\subsection{Database Comparison (DBCOMP)%
  \label{database-comparison-dbcomp}%
}

The last verification step is to substantiate the main aim of this project, generating layouts with Python (PyCells) precisely similar to those generated with SKILL (PCells). This verification process is in GDSII level and is performed with the help of a regression test, by which the PCells and PyCells of identical devices with a same set of parameters are instanced in a layout, and then these instances are compared to be congruent.

\section{Conclusion%
  \label{conclusion}%
}

Believing in what Richard Stallman has said \textquotedbl{}Proprietary software is an injustice\textquotedbl{}, we have successfully developed the layout generators for an open semiconductor industry, giving the designers who use ams AG technologies, opportunity to work with any desired EDA-tools, to share their designs easily with others, and to receive the support they deserve. These PyCells are developed for all high and low voltage devices of ams AG technologies in order to enable the IC-designers to have advanced designs without having constraints because of missing parametrized layouts. Furthermore, by using these PyCells we have managed to reduce the computation time which is extremely important in semiconductor industry. The developed PyCells have passed all the verification tests and in practical use by customers. The PyCells are available free of charge to all Europractice \cite{Europractice} members, currently more than 500 universities and more than 100 research institutes worldwide active in IC design are members of Europractice, and have free access to the technologies of ams AG including the PyCells.

\end{document}